\documentclass[aps,prb,twocolumn,showpacs]{revtex4}
\usepackage[dvips]{graphicx}
\usepackage{amssymb}

\begin{document}

\title{Dynamical Monte Carlo investigation of spin reversals and nonequilibrium magnetization of single-molecule magnets}

\author{Gui-Bin Liu}
\author{Bang-Gui Liu}
\email[Corresponding author:~]{bgliu@mail.iphy.ac.cn}
\affiliation{Institute of Physics, Chinese Academy of Sciences, Beijing 100190, China\\
Beijing National Laboratory for Condensed Matter Physics, Beijing 100190, China}

\date{\today}

\begin{abstract}
In this paper, we combine thermal effects with Landau-Zener (LZ)
quantum tunneling effects in a dynamical Monte Carlo (DMC) framework
to produce satisfactory magnetization curves of single-molecule
magnet (SMM) systems. We use the giant spin approximation for SMM
spins and consider regular lattices of SMMs with magnetic dipolar
interactions (MDI). We calculate spin reversal probabilities from
thermal-activated barrier hurdling, direct LZ tunneling, and
thermal-assisted LZ tunnelings in the presence of sweeping magnetic
fields. We do systematical DMC simulations for Mn$_{12}$ systems
with various temperatures and sweeping rates. Our simulations
produce clear step structures in low-temperature magnetization
curves, and our results show that the thermally activated barrier
hurdling becomes dominating at high temperature near 3K and the
thermal-assisted tunnelings play important roles at intermediate
temperature. These are consistent with corresponding experimental
results on good Mn$_{12}$ samples (with less disorders) in the
presence of little misalignments between the easy axis and applied
magnetic fields, and therefore our magnetization curves are
satisfactory. Furthermore, our DMC results show that the MDI, with
the thermal effects, have important effects on the LZ tunneling
processes, but both the MDI and the LZ tunneling give place to the
thermal-activated barrier hurdling effect in determining the
magnetization curves when the temperature is near 3K. This DMC
approach can be applicable to other SMM systems, and could be used
to study other properties of SMM systems.

\end{abstract}


\maketitle

\section{Introduction}\label{sec1}

Single-molecule magnet (SMM) systems attract more and more
attention because they can be used to make devices for spintronic
applications \cite{spintr1,spintr2}, quantum computing \cite{qc},
high-density magnetic information storage \cite{mem} etc
\cite{ad21,bb,ad22}. Usually, a SMM can be treated as a large spin
with strong magnetic anisotropy at low temperature. The most
famous and typical is Mn$_{12}$-ac
([Mn$_{12}$O$_{12}$(Ac)$_{16}$(H$_2$O)$_4$]$\cdot$2HAc$\cdot$4H$_2$O,
where HAc=accetic acid), or Mn$_{12}$ for short \cite{Mn12a}. It
usually has spin $S=10$ and large anisotropy energy, producing a
high spin reversal barrier \cite{sessoli}. Many interesting
phenomena have been observed, such as various dynamical magnetism.
One of the most intriguing phenomena observed in SMM systems is a
step-wise structure in low-temperature magnetization curves
\cite{mstep1,mstep2,mstep3}. Great efforts have been made to
investigate this phenomenon and related effects
\cite{add1,chio,Mn12tBuAc,TB,Mn4b,book,Mn12para1,Mn12E}. The
step-wise structure is attributed to Landau-Zener (LZ) quantum
tunneling effect \cite{Landau,Zener}. This stimulates intensive
study on LZ model and its variants
\cite{lz1,lz2,lz3,lzepl,lz4,lz5,lz6,lz7,lz8}. Some authors use
numeric diagonalization methods \cite{numeric1,numeric2} to study
many-level LZ models to understand the step structure in
experimental magnetization curves. However, it is difficult to
consider thermal effects in these approaches to obtain
satisfactory magnetization curves comparable to experimental
results.

In this paper, we shall combine the classical thermal effects with
the quantum LZ tunneling effects in a dynamical Monte Carlo (DMC)
framework \cite{dmc0,dmc1,gbliu} in order to produce satisfactory
magnetization curves comparable to experimental results. We consider
ideal tetragonal body-centered lattices and use the giant spin
approximation for spins of SMMs. We consider magnetic dipolar
interactions, but neglect other factors such as defects, disorders,
and misalignments between the easy axis and applied magnetic field.
We calculate spin reversal probabilities from thermal-activated
barrier hurdling, direct LZ tunneling effect, and thermal-assisted
LZ tunneling effects in the presence of sweeping magnetic fields,
and thereby derive a unified probability expression for any
temperature and any sweeping field. Taking the Mn$_{12}$ as example,
we do systematical DMC simulations with various temperatures and
sweeping rates. The step structure appears in our simulated
low-temperature magnetization curves, and our simulated
magnetization curves are semi-quantitatively consistent with
corresponding experimental results on those good Mn$_{12}$ systems
(with less disorders) in the presence of little misalignments
between the easy axis and applied fields \cite{Mn12tBuAc,TB}.
Interplays of the LZ tunneling effect, the thermal effects, and the
magnetic dipolar interactions are elucidated. These imply that our
simple model and DMC method capture the main features of
experimental magnetization curves for little misalignments. More
detailed results will be presented in the following.

The rest of this paper is organized as follows. In next section we
shall define our spin model and describe approximation strategy.
In the third section we shall describe our simulation method,
present our unified probability formula for the spin reversal from
the three spin reversal mechanisms, and give our simulation
parameters. In the fourth section we shall present our simulated
magnetization curves and some analysis. In the fifth section we
shall show the key roles of the dipolar interactions in
determining LZ tunneling probabilities. Finally, we shall give our
conclusion in the sixth section.

\section{Spin Model and approximation}

Without losing generality, we take typical Mn$_{12}$ system as our
sample in the following. Under giant spin approximation, every
Mn$_{12}$ SMM is represented by a spin $S$=10. Magnetic dipolar
interactions are the only inter-SMM interactions, with hyperfine
interactions neglected. Mn$_{12}$ SMMs are arranged to form a
body-centered tetragonal lattice with experimental lattice
parameters\cite{Mn12abc}. Using a body-centered tetragonal unit
cell that consists of two SMMs, we define our lattice as
$L_1\times L_2\times L_3$, where $L_1$, $L_2$, and $L_3$ are three
positive integers. A longitudinal magnetic field $B_z(t)=B_0+\nu
t$ is applied along the $c$-easy axis of magnetization, where
$\nu$ is the field-sweeping rate and $B_0$ is the starting
magnetic field. The total Hamiltonian of this system can be
expressed as
\begin{equation}
\hat{H}=\sum_i\hat{H}_i^0+\frac12\sum_{i\ne j}\hat{H}_{ij}^{\rm
di}~, \label{eq.Htot}
\end{equation}
where $\hat{H}_i^0$ is the single-body part for the $i$-th single
SMM, and $\hat{H}_{ij}^{\rm di}$ describes the magnetic dipolar
interaction between the $i$-th and $j$-th SMM. The factor $1/2$
before the sum sign is due to the double counting in the
summation. $\hat{H}_i^0$ is given by
\begin{eqnarray}
\hat{H}_i^0 = & -D(\hat{S}_i^z)^2 + E[(\hat{S}_i^x)^2-(\hat{S}_i^y)^2] \nonumber\\
 & + B_4^0\hat{O}_4^0+B_4^4\hat{O}_4^4 + g\mu_B B_z \hat{S}_i^z ~,
\label{eq.Hi0}
\end{eqnarray}
where
$\hat{\mathbf{S}}_i\equiv(\hat{S}_i^x,\hat{S}_i^y,\hat{S}_i^z)$ is
the spin vector operator for the $i$-th SMM, $g$ the Land\'{e}
g-factor (here $g=2$ is used), $\mu_B$ the Bohr magneton, $D$,
$E$, $B_4^0$ and $B_4^4$ are all anisotropic parameters, and
$\hat{O}_4^0$ and $\hat{O}_4^4$ are both Steven
operators\cite{book} defined by
$\hat{O}_4^0=35(\hat{S}_i^z)^4-[30S(S+1)-25](\hat{S}_i^z)^2+3S^2(S+1)^2-6S(S+1)$
and $\hat{O}_4^4=[(\hat{S}_i^+)^4+(\hat{S}_i^-)^4]/2$.
$\hat{H}_{ij}^{\rm di}$ is defined by
\begin{equation}
\hat{H}_{ij}^{\rm
di}=\frac{\mu_0g^2\mu^2_B}{4\pi r^3_{ij}}[\hat{\mathbf{S}}_i\cdot\hat{\mathbf{S}}_j-\frac{3}{r^2_{ij}}(\hat{\mathbf{S}}_i\cdot
\mathbf{r}_{ij})(\hat{\mathbf{S}}_j\cdot \mathbf{r}_{ij})]~,
\label{eq.Hijdi}
\end{equation}
where $\mu_0$ is the magnetic permeability of vacuum, and
$\mathbf{r}_{ij}$ the vector from $i$ to $j$, with $r_{ij}$=$|
\mathbf{r}_{ij}|$ being the distance between $i$ and $j$.

For the $i$-th SMM, we treat all the effects from the other SMMs
by classical-spin approximation. As a result, we derive the
partial Hamiltonian $\hat{H}_i$ that acts on the $i$-th SMM:
\begin{eqnarray}
\! \hat{H}_i &\!\!=\!\!& \hat{H}_i^0+g\mu_B \mathbf{B}_i^{\rm di}\cdot \hat{\mathbf{S}}_i\nonumber\\
&\!\!=\!\!&  -D(\hat{S}_i^z)^2+B_4^0\hat{O}_4^0+\hat{H}_i^{\rm tr}
+ g\mu_B (B_z + B_{iz}^{\rm di})\hat{S}_i^z, \label{eq.Hi}
\end{eqnarray}
where the transverse part $\hat{H}_i^{\rm tr}$ is defined as
\begin{eqnarray}
\hat{H}_i^{\rm tr} = E[(\hat{S}_i^x)^2\!-\!(\hat{S}_i^y)^2] +
B_4^4\hat{O}_4^4 + g\mu_B(B_{ix}^{\rm
di}\hat{S}_i^x\!+\!B_{iy}^{\rm di}\hat{S}_i^y). \label{tran}
\end{eqnarray}
For the $i$-th SMM, the dipolar interaction of the other SMMs is
equivalent to $\mathbf{B}_i^{\rm di}\equiv(B_{ix}^{\rm di},
B_{iy}^{\rm di}, B_{iz}^{\rm di})= \sum_{j(\ne
i)}\mathbf{B}_{ji}^{\rm}$, where $\mathbf{B}_{ji}^{\rm}$ is the
magnetic dipolar field applied by the $j$-th SMM on the $i$-th
SMM. It contributes a magnetic field consisting of longitudinal
and transverse parts.

\begin{figure*}[!htbp]
\includegraphics[width=13cm]{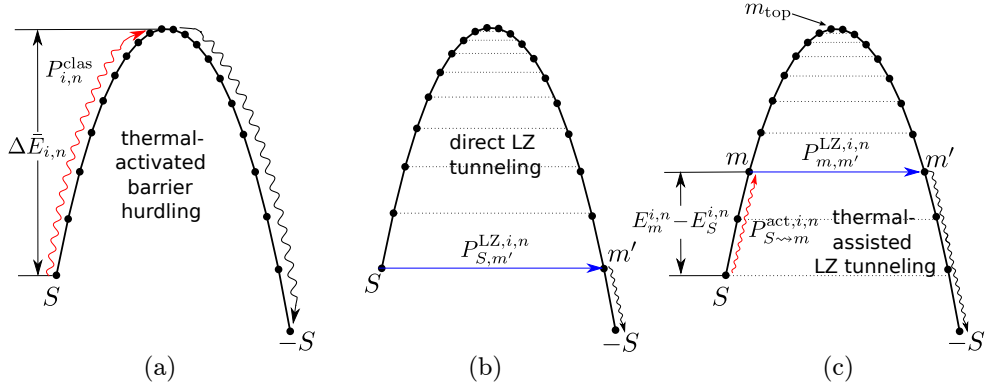}
\caption{(Color online.) A schematic demonstration of the three
spin reversal mechanisms: (a) thermal-activated barrier hurdling,
(b) direct LZ tunneling, and (c) thermal-assisted LZ tunneling.
The probabilities, energy levels, barrier, and other symbols are
defined in the text. The horizontal solid line with arrow in (b)
and (c) shows that a pair of energy levels satisfy the resonance
tunneling conditions. The horizontal dotted lines in (b) and (c),
as guide for eyes, imply that these energy levels do not match. }
\label{fig:1}
\end{figure*}

\section{Simulation method and parameters}

As we show in Fig. 1, there are three main mechanisms related to
the reversal of a SMM spin
\cite{sessoli,book,mstep1,mstep2,mstep3,add1,lz1,lzepl}: (a)
thermal-activated barrier-hurdling, (b) direct LZ tunneling, and
(c) thermal-assisted LZ tunneling. The thermal-activated barrier
hurdling dominates at high temperature (if the blocking
temperature $T_B\sim$3.3K for Mn$_{12}$ \cite{TB} is treated as
high temperature), the direct LZ tunneling at low temperature, and
the thermal-assisted LZ tunneling at intermediate temperature. For
any temperature, we consider all the three spin reversal
mechanisms simultaneously. For the time scale we are interested,
we do not need to treat phonon-related interactions directly, but
shall use an effective transition-state theory to calculate the
thermal-activated spin-reversal rates. We shall use a DMC method
to combine the quantum LZ tunneling effects with the classical
thermal effects. Various kinetic Monte Carlo (KMC)
methods\cite{witten,kmc1,kmc2,lbg99,kmc3}, essentially similar to
this DMC method, have been used to simulate atomic kinetics during
epitaxial growth for many years. On the other hand, MC simulation
has been used to study Glauber dynamics of kinetic Ising
models\cite{glauber,ad23,ad24}. We shall present a detailed
description of this DMC simulation method in the following.

\subsection{Thermal-activated spin reversal probability}

We need the thermal-activated energy barrier in order to calculate
the thermal-activated spin-reversal rate. When calculating the
thermal-activated energy barrier we ignore the small transverse
part $\hat{H}_i^{\rm tr}$ and use classical approximation for the
spin operators. The large spin $S=10$ of Mn$_{12}$ further
supports the approximations. As a result, the energy of the $i$-th
SMM can be expressed as
\begin{equation}
\bar{E}_i=-D_2(S_i^z)^2-D_4(S_i^z)^4+h_iS_i^z~, \label{eq.Ei}
\end{equation}
where $S_i^z$ is the classical variable for the spin operator
$\hat{S}_i^z$, $h_i=g\mu_B (B_z+B_{iz}^{\rm di})$,
$D_2=D+[30S(S+1)-25]B_4^0$, and $D_4=-35B_4^0$. Because $h_i$ is
dependent on time $t$, $\bar{E}_i$ changes with $t$.

We define our MC steps by the time points, $t_n=\Delta t\cdot n$,
where $n$ takes nonnegative integers in sequence. For the $n$-th
MC step, we use $\bar{E}_{i,n}$, $h_{i,n}$, and $S_{i,n}^z$ to
replace $\bar{E}_i$, $h_i$, and $S_i^z$. Because each of the spins
has two equilibrium orientations along the easy axis, we assume
every spin takes either $S$ or $-S$ at each of the times $t_n$.
Within the $n$-th MC step ($t$: from $t_n$ to $t_{n+1}$), we use
an angle variable $\theta_{i,n}$ to describe the $i$-th spin's
deviation from its original ($t_n$) orientation $S_{i,n}^{\rm
eq}$. Naturally, $\theta_{i,n}=0$ corresponds to the original
state and $\theta_{i,n}=\pi$ is the reversed state, and then all
the other angle values ($0<\theta_{i,n}<\pi$) are treated as
transition states. Expressing $S_{i,n}^z$ as $S_{i,n}^{\rm
eq}\cos\theta_{i,n}$, we usually have a maximum in the curve of
$\bar{E}_{i,n}(\cos\theta_{i,n})$ as a function of
$\cos\theta_{i,n}$, and the maximum determines the energy barrier
for the spin reversal mechanism\cite{liying1,liying2,lbg}, as
shown in Fig. 1(a). We define $x_{i,n}=\cos\theta_{i,n}$ for
convenience. We have $-1\leq x_{i,n} \leq 1$ for actual
$\theta_{i,n}$, but $x_{i,n}$ can be extended beyond this region
in order to always obtain a formal solution $x_{i,n}^{\rm max}$
for the maximum. $|x_{i,n}^{\rm max}|< 1$ implies that there
actually exists an energy barrier, and $|x_{i,n}^{\rm max}|\ge 1$
means that there is no barrier for the corresponding process.
Under conditions $D_2>0$ and $D_4>0$, the barrier can be expressed
as:
\begin{equation}
\Delta \bar{E}_{i,n} = \left\{ \begin{array}{lll}
\bar{E}_{i,n}(x_{i,n}^{\rm max}), && |x_{i,n}^{\rm max}|\leq
1\\
\bar{E}_{i,n}(-1)=|2h_{i,n}S_{i,n}^{\rm eq}|,  && x_{i,n}^{\rm max}<-1\\
\bar{E}_{i,n}(1)=0,  && x_{i,n}^{\rm max}>1
\end{array} \right.
\end{equation}
where $x_{i,n}^{\rm max}$ is defined by
\begin{equation}
x_{i,n}^{\rm max}=\sqrt[3]{-q_{i,n}/2+\sqrt{d_{i,n}}} +
\sqrt[3]{-q_{i,n}/2-\sqrt{d_{i,n}}}
\end{equation}
and the three parameters are defined by
$d_{i,n}=(q_{i,n}/2)^2+(p/3)^3$, $p=D_2/(2D_4S^2)$, and
$q_{i,n}=-h_{i,n}S_{i,n}^{\rm eq}/(4D_4S^4)$. These parameters are
dependent on the spin configuration and the magnetic field, and
then on the time $t_n$ (or $n$).

The spin reversal rate within the $n$-th MC step (between $t_n$
and $t_{n+1}$) can be expressed as $R_{i,n}=R_0\exp(-\Delta
\bar{E}_{i,n}/k_B T)$ in terms of Arrhenius law
\cite{ArrheniusLaw2}, where $k_B$ is the Boltzmann constant and
$R_0$ the characteristic attempt frequency.  We use $P_n(t')$ to
describe the probability that the $i$-th spin is reversed between
$0$ and $t'$, where $t'$ satisfies the condition $t'\le\Delta t$.
It has the initial condition $P_n(t'=0)=0$ and satisfies the
equation $[1-P_n(t')]\cdot R_n(t')dt'=P_n(t'+dt')-P_n(t')$, or
\begin{equation}
 [1-P_n(t')]R_n(t')=\frac{d}{dt'}P_n(t')
\end{equation}
where $R_n(t')$, the reversal rate at $t'$, is taken as the rate
$R_{i,n}$, independent of $t'$ within the region [0,$\Delta t$].
Solving the equation, we obtain the probability $P^{\rm
clas}_{i,n}$ defined as $P_n(t'=\Delta t)$ for a classical
thermal-activated reversal of the $i$-th spin within the $n$-th MC
step:
\begin{equation}
\label{Pclas}
   P^{\rm clas}_{i,n}=1-\exp(-\Delta t\cdot R_{i,n}).
\end{equation}
For $\Delta t\ll 1/R_{i,n}$, Eq. (\ref{Pclas}) reduces to $P^{\rm
clas}_{i,n}=\Delta t\cdot R_{i,n}$. The probability expression
defined in Eq. (\ref{Pclas}) is reasonable because $P^{\rm
clas}_{i,n}$ will not exceed unity even when $\Delta t$ is very
large with respect to $1/R_{i,n}$.

\subsection{LZ-tunneling related spin reversal probabilities}

When temperature is lower than $T_B$, LZ tunneling begins to
contribute to spin reversal. We begin with the effective quantum
single-spin Hamiltonian (\ref{eq.Hi}) with (\ref{tran}). All the
effects of other spins are included in the magnetic dipolar field
$\mathbf{B}_{i}^{\rm di}$ (depending on the time $t$) and are
depending on the magnetic field and the current spin
configuration. For the $n$-th MC step, if the transverse term
$\hat{H}_i^{\rm tr}$ is removed, Hamiltonian Eq. (\ref{eq.Hi}) is
diagonal and has $2S+1$ energy levels, $E_{m}^{i,n}$, where $m$
can take any of $S, S-1,\cdots,-(S-1),-S$. If using the continuous
time variable $t$, we can express the energy levels as
$E_{m}^{i}(t)$ (with $m$ from $S$ to -$S$) and derive their
crossing fields [at which $E_{m}^{i}(t)=E_{m'}^{i}(t)$]:
\begin{equation}
 B_{m,m'}=\frac{(m+m')[D_2+D_4(m^2+m'^2)]}{g\mu_B}.
\end{equation}
The transverse term $\hat{H}_i^{\rm tr}$ will modify the energy
levels $E_{m}^{i,n}$, but the $2S+1$ energy levels of Hamiltonian
(\ref{eq.Hi}) with (\ref{tran}), $\tilde{E}_{m}^{i,n}$, can be
still labelled by $m=S, S-1,\cdots,-(S-1),-S$. Actually, the
difference between $E_{m}^{i,n}$ and $\tilde{E}_{m}^{i,n}$ is
small. Due to the existence of the transverse part $\hat{H}_i^{\rm
tr}$, there will be an avoided level crossing between
$\tilde{E}_{m}^{i,n}$ and $\tilde{E}_{m'}^{i,n}$ for the $n$-th MC
step when the effective field $B_z+B_{iz}^{\rm di}$ equals
$B^{i,n}_{m,m'}$, with $m$ and $m'$ taking values among $S,
S-1,\cdots,-(S-1),-S$. The set of all the $B^{i,n}_{m,m'}$ values
are the effective field conditions for the
avoided-level-crossings. If $E_{m}^{i,n}$ equals $E_{m'}^{i,n}$,
$B^{i,n}_{m,m'}$ is approximately equivalent to the crossing field
(equaling $B_{m,m'}$). The allowed $(m,m')$ pairs are shown in
Fig. 2. This means that when $B_z$ is swept to a right
$B_{m,m'}^{i,n}-B_{iz}^{\rm di}$ value, a quantum tunneling occurs
between the $m$ and $m'$ states. The tunneling can be well
described using LZ tunneling\cite{gbliu,Mn4b,numeric1,numeric2}.
The nonadiabatic LZ tunneling probability $P^{{\rm
LZ},i,n}_{m,m'}$ is given by\cite{Landau,Zener}
\begin{equation}
P^{{\rm LZ},i,n}_{m,m'}=1-\exp \Bigg[ -\frac{\pi
(\Delta_{m,m'}^{i,n})^2}{2\hbar g\mu_B|m-m'|\nu} \Bigg] ~,
\label{Plz}
\end{equation}
where the tunnel splitting $\Delta^{i,n}_{m,m'}$ is the energy gap
at the avoided crossing of states $m$ and $m'$. $B^{i,n}_{m,m'}$
and $\Delta^{i,n}_{m,m'}$ can be calculated by diagonalizing Eq.
(\ref{eq.Hi}). If the dipolar field is neglected,
$B^{i,n}_{m,m'}$, $\Delta^{i,n}_{m,m'}$, and $P^{{\rm
LZ},i,n}_{m,m'}$ reduce to $B^0_{m,m'}$, $\Delta^0_{m,m'}$, and
$P^0_{m,m'}$, those of corresponding isolated SMMs, respectively.

\begin{figure*}[!htbp]
\includegraphics[width=14cm]{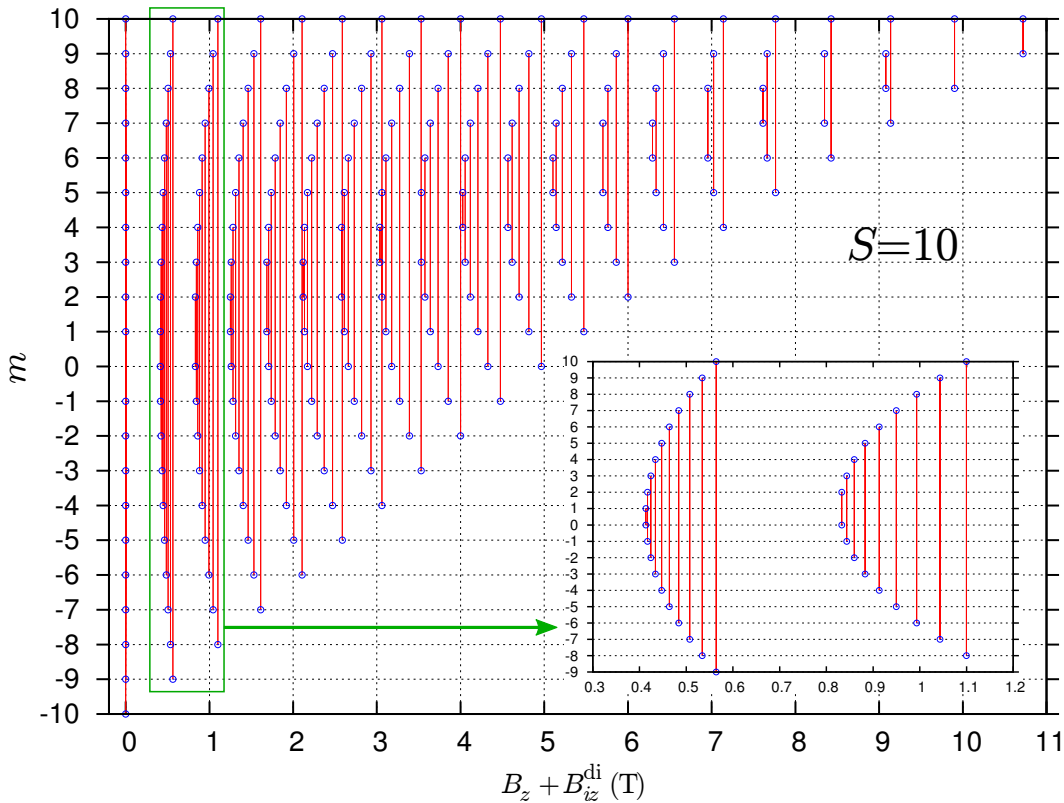}
\caption{(Color online.) A schematic demonstration for the
conditions of the $i$-th spin that Landau-Zener tunnelings can
happen. $m$ labels the spin $z$ component, from -10 to 10, and
$B_z+B_{iz}^{\rm di}$ is the effective magnetic field. A hollow
circle indicates one of the allowed $m$ values. The two circles
($m$ and $m'$) connected with one vertical line means that a LZ
tunneling condition is satisfied between the two states at the
corresponding effective field $B^{i,n}_{m,m'}$ within the $n$-th
MC step. There is at most one LZ tunneling at any nonzero value of
the effective field, but  every $m$ state can tunnel to the -$m$
state when the effective field is zero. The inset amplifies the
part between 0.3 and 1.2 T.} \label{fig:2}
\end{figure*}

At the beginning of field sweeping, we let all the spins have
$m=S$. If $T\!\ll\!T_B$, thermal activations are frozen, and LZ
tunnelings only occur at the avoided crossings $(S,m')$, where
$m'$ takes one of $-S,-S+1,\cdots,S-1$. This is the direct
tunneling shown in Fig. 1(b), and the LZ tunneling probability is
given by
\begin{equation}
 P^{\rm dLZ}_{i,n}=P_{S,m'}^{{\rm dir},i,n}=P_{S,m'}^{{\rm
 LZ},i,n}.\label{Pdlz}
\end{equation}
It is nonzero only when the condition $E_{S}^{i,n}=E_{m'}^{i,n}$
is satisfied. When the temperature is in the intermediate region
$0\ll T<T_B$, the thermal-assisted tunneling plays an important
role. This process can be represented by $S\rightsquigarrow m
\rightarrow m'$ as shown in Fig. 1(c), in which $S$ and $m$ states
lie on one side of the thermal barrier and $m'$ and $-S$ states on
the other side. The first process $S\rightsquigarrow m$ means that
a spin is thermally activated from $S$ to $m$ state with the
probability $P_{S\rightsquigarrow m}^{\text{act},i,n}$, which is
given by $P_{S\rightsquigarrow m}^{\text{act},i,n}=1-\exp(-\Delta
t\cdot R^{\text{act}}_{i,n})$, where $R^{\text{act}}_{i,n}$ is
given by $R_0\exp[-(E_{m}^{i,n}-E_{S}^{i,n})/k_B T]$.  The second
process $m\rightarrow m'$ is the LZ tunneling from $m$ to $m'$,
with the probability defined in Eq.~(\ref{Plz}). Therefore, the
reversal probability of thermal-assisted LZ tunneling through $m$
is given by
\begin{equation}P^{\rm
taLZ}_{i,n,m}=P_{S\rightsquigarrow m\rightarrow
m'}^{\text{ass},i,n}=P_{S\rightsquigarrow
m}^{\text{act},i,n}P_{m,m'}^{{\rm LZ},i,n}.\label{Ptalz}
\end{equation}
It is nonzero only when the condition $E_{m}^{i,n}=E_{m'}^{i,n}$
is satisfied.

It must be pointed out that $m'$ in $P_{S,m'}^{{\rm dir},i,n}$ and
$P_{S\rightsquigarrow m\rightarrow m'}^{\text{ass},i,n}$ is
determined by $E_{m'}^{i,n}=E_{S}^{i,n}$ and
$E_{m'}^{i,n}=E_{m}^{i,n}$, respectively, as is shown in Figs.
1(b) and 1(c). If the energy-level condition is satisfied, the
probability is larger than zero; or else the probability is
equivalent to zero.
Therefore, the subscript $m'$ in $P_{S,m'}^{{\rm dir},i,n}$ and
$P_{S\rightsquigarrow m\rightarrow m'}^{\text{ass},i,n}$ can be
removed, as we have done in $P^{\rm dLZ}_{i,n}$ and $P^{\rm
taLZ}_{i,n,m}$.

\subsection{Unified spin reversal probability for MC simulation}

Generally speaking, every one of the three spin reversal
mechanisms takes action at any given temperature. Actually the LZ
tunneling effect dominates at low temperatures and the thermal
effects become more important at higher temperatures. For the
$n$-th MC step, the probability for the thermal-activated
barrier-hurdling reversal of the $i$-th spin is given by $P^{\rm
clas}_{i,n}$ defined in Eq. (\ref{Pclas}) [see Fig. 1(a)], that
for the direct LZ tunneling effect equals $P^{\rm dLZ}_{i,n}$
defined in Eq. (\ref{Pdlz}) [see Fig. 1(b)], and that for the
thermal-assisted LZ tunneling effects through the $m$ state is
given by $P^{\rm taLZ}_{i,n,m}$  defined in Eq. (\ref{Ptalz}) [see
Fig. 1(c)]. Here the partial probabilities from the three
mechanisms are considered independent of each other. Therefore, we
can derive the total probability $P_{i,n}^{\rm tot}$ for the
reversal of the $i$-th spin within the $n$-th MC step:
\begin{equation}
P_{i,n}^{\rm tot}=1-(1-P_{i,n}^{\rm clas})(1-P^{\rm dLZ}_{i,n})
\!\!\prod_{m_{\rm top}<m<S}\!\!(1-P^{\rm taLZ}_{i,n,m}),
\label{Ptot}
\end{equation}
where $m_{\rm top}$, depending on the effective field, is
determined by the highest level $E^{i,n}_{m_{\rm top}}$ among the
$2S$ energy levels, $E^{i,n}_{m}$ ($-S\le m <S$), as we show in
Fig. 1(c).

It must be pointed out that $P_{i,n}^{\rm clas}$ is always larger
than zero, but the LZ-tunneling related probabilities, $P^{\rm
dLZ}_{i,n}$ and $P^{\rm taLZ}_{i,n,m}$, are nonzero only at some
special values of the effective field. As is shown in Fig. 2,
there is at most one LZ-tunneling channel, from either direct or
thermal-assisted LZ effect, for a given nonzero value of the
effective field. As a result, when the effective field is nonzero,
we have at most one nonzero value from either $P^{\rm dLZ}_{i,n}$
or one of $P^{\rm taLZ}_{i,n,m}$ ($m_{\rm top}<m<S$). It is only
at the zero value of the effective field that both $P^{\rm
dLZ}_{i,n}$ and $P^{\rm taLZ}_{i,n,m}$ ($0<m<S$) ($m_{\rm top}=0$)
can be larger than zero so that we can have the direct LZ
tunneling and all the thermal-assisted LZ-tunneling channels
simultaneously. In our simulations, the processes that a reversed
spin is reversed again are also considered, but the probabilities
are tiny.

\subsection{Simulation parameters}

We use experimental lattice constants, $a=b=17.1668\,\text{\AA}$
and $c=12.2545\, \text{\AA}$, and experimental anisotropy
parameters, $D/k_B=0.66$\,K, $B_4^0/k_B=-3.2\times10^{-5}$\,K, and
$B_4^4/k_B=6\times10^{-5}$\,K \cite{Mn12abc,Mn12tBuAc,Mn12para1}.
As for the second-order transverse parameter $E$,
$E/k_B=1.8\times10^{-3}$\,K is taken from the average of
experimental values \cite{Mn12E}. We describe the time by using
both continuous variable $t$ and discrete superscript/subscript
$n$. In some cases, the sweeping field can be used to describe the
time because it is defined by $B_z(t)=B_0+\nu t$. There is always
a nonnegative integer $n$ for any given $t$ value, and there is a
$t$ region, [$t_n,t_{n+1}$], for any given nonnegative $n$. We
take $\Delta t=0.1$\,ms and $R_0=10^9$/s, which guarantee the good
balance between computational demand and precision.

The dipolar fields $(B_{ix}^{\rm di},B_{iy}^{\rm di},B_{iz}^{\rm
di})$ at each SMM are updated whenever any of the SMM spins is
reversed. The $\Delta^{i,n}_{m,m'}$ values are recalculated
whenever any LZ tunneling happens. In the simulations, the field
$B_z$ is swept from -7 to 7\,T in the forward process, and the
full magnetization hysteresis loop is obtained simply by using the
loop symmetry. Every magnetization curve is calculated by
averaging over 100 runs to make statistical errors small enough.
The main results presented in the following are simulated and
calculated with lattices consisting approximately of $900\sim
1200$ body-centered unit cells or $1800\sim 2400$ spins. We have
test our results with lattices consisting approximately of 100
$\sim$ 6000 body-centered unit cells, or 200 $\sim$ 12000 spins.

\section{simulated magnetization curves}

Presented in Fig. 3 are simulated magnetization curves (with $M$
normalized to the saturated value $M_S$) against the applied
sweeping field $B_z$ for ten different temperatures: 0.1, 0.5,
0.6, 0.8, 1.0, 1.5, 2.0, 2.5, 2.8, and 3.2 K. Here, the lattice
dimension is $10\times10\times10$ and the field sweeping rate is
0.02\,T/s. Each of the curves is calculated by averaging over 100
runs. The curves of 0.1\,K and 0.5\,K fall in the same curve,
which implies that thermal activation is totally frozen when the
temperature is below 0.5\,K. It can be seen in Fig. 3 that the
area enclosed by a magnetization loop decreases with the
temperature increasing, becoming nearly zero at 3.2 K (near the
blocking temperature 3.3\,K of Mn$_{12}$). There are clear
magnetization steps when the temperature is below 2.0\,K. They are
caused by the LZ quantum tunneling effects. For convenience, we
describe a step by using a H-part, a vertex, and a V-part. For an
ideal step, the H-part is horizontal and the V-part vertical, but
for any actual step in a magnetization curve, the H-part is not
horizontal and the V-part not vertical because of the dipolar
interaction and thermal effects, and the two parts still meet at
the vertex. The vertex is convex toward the up-left direction in
the right part of a magnetization loop and toward the down-right
direction in the left part. At higher temperatures ($\geq 2.0$K),
there is no complete step and there are only some kinks that
remind us of some LZ tunnelings. This should be caused mainly by
thermal effects.

\begin{figure}[!htbp]
\includegraphics[width=8.5cm]{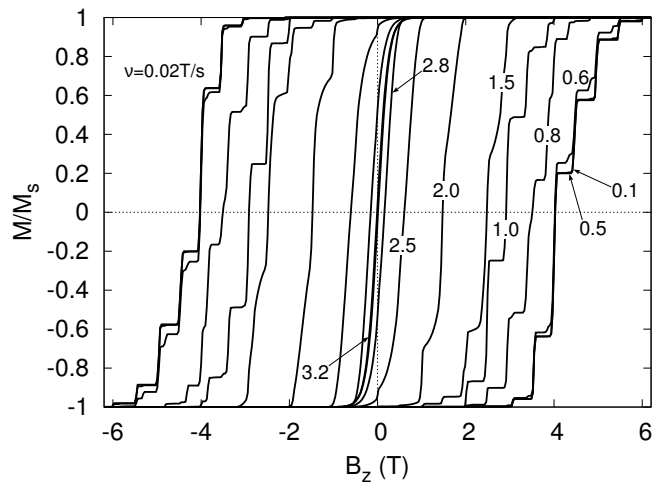}
\caption{Simulated magnetic hysteresis loops ($M/M_S$ vs $B_z$)
with sweeping rate $\nu=0.02$\,T/s for ten temperatures: 0.1, 0.5,
0.6, 0.8, 1.0, 1.5, 2.0, 2.5, 2.8, and 3.2\,K (from outside to
inside). The lattice dimension is $10\times10\times10$. Note that
the two curves of 0.1\,K and 0.5\,K fall in the same curve. }
\label{fig:3}
\end{figure}

Presented in Fig. 4  are the right parts of the magnetization
curves against the applied sweeping field for three temperatures,
0.1, 1.5, and 2.5 K, and with three sweeping rates, 0.002, 0.02,
and 0.2 T/s. Here, the lattice dimension is $10\times10\times10$.
We label a magnetization step by the magnetic field defined by its
V-part near its vertex. For $T$=0.1K, only the direct LZ
tunnelings change the magnetization, and the magnetization steps
from $B_z$=2 to 6T in Fig. 4 correspond to $B_{S,m'}^0$ with $m'$
being from -6 to 2 in Table I. For $T$=1.5K, there are clear steps
in the lower parts of the three magnetization curves, but their
V-parts deviate substantially from the corresponding $B_{S,m'}^0$
values and the steps are substantially deformed, which show that
thermal-assisted LZ tunnelings play an important role. When
temperature rises to 2.5K, there is no step structure and only one
kink can be seen in the lower part of the magnetization curve in
the cases of 0.2T/s and 0.02T/s. This is because the effects of
thermal activation become dominating over the LZ tunneling
effects. Different sweeping rates lead to substantial changes in
the magnetization curves, and the larger the sweeping rate
becomes, the larger the hysteresis loops are.

\begin{figure}[!htbp]
\includegraphics[width=8.2cm]{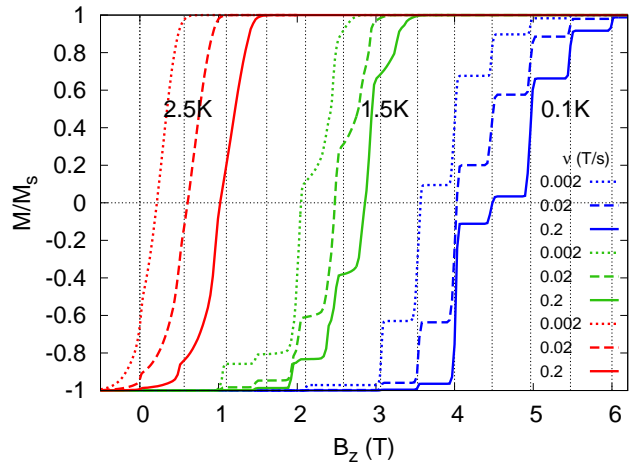}
\caption{(Color online.) The right parts of simulated
magnetization curves of different sweeping rates 0.002 (dot), 0.02
(dash), and 0.2 (solid)\,T/s for three temperatures 0.1\,K,
1.5\,K, and 2.5\,K, as labelled. The lattice dimension is
$10\times10\times10$. Each of the visible steps and kinks along a
magnetization curve corresponds to one of the magnetic fields at
which the direct and thermal assisted LZ tunnelings take place.
The thin vertical dotted lines show the positions of $B_{S,m'}^0$
for $m'$=-10,-9,$\cdots$,2.} \label{fig:4}
\end{figure}

\begin{figure}[!htbp]
\includegraphics[width=8.3cm]{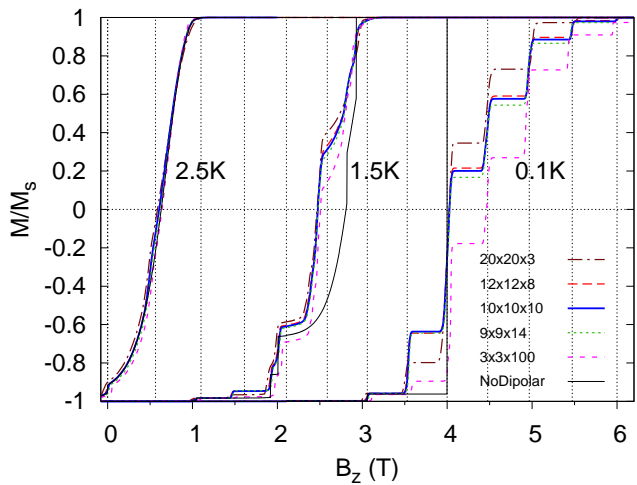}
\caption{(Color online.) The right parts of simulated
magnetization curves for three temperatures with five different
lattice dimensions: $20\times20\times3$ (dash-dot),
$12\times12\times8$ (dash), $10\times10\times10$ (solid),
$9\times9\times14$ (dot), and $3\times3\times100$ (short-dash).
The temperatures are 0.1\,K, 1.5\,K, and 2.5\,K, as labelled. The
sweeping rate is 0.02\,T/s. For comparison, we also present the
results without considering dipolar interaction (thin solid line).
} \label{fig:5}
\end{figure}

Presented in Fig. 5 are the right parts of simulated hysteresis
loops with $\nu$=0.02 T/s at three temperatures for five different
lattice dimensions: $20\times20\times3$, $12\times12\times8$,
$10\times10\times10$, $9\times9\times14$, and $3\times3\times100$.
The temperatures are 0.1, 1.5, and 2.5\,K. For comparison, the
simulated results without the dipolar interaction are presented
too. For $T$=0.1K, there are clear step structures for all the
five lattice shapes. The step height varies with the lattice
shape, which can be attributed to the dipolar interaction. If the
dipolar interaction is switched off, there are only two steps: one
tall step at $B_z=4.00$ T and the other very short step at 3.06 T.
They correspond to the two transitions from 10 to -2 and -4,
respectively. Other transitions from 10 to -4, -6, -8, and -10
have too small probabilities to be seen. When the dipolar
interaction is switched on, the transition from 10 to -3 is
allowed and the tall step becomes much shorter, resulting in the
rich step structures between 3 and 6 T. The steps are caused by
the direct LZ tunnelings. When the temperature changes to 1.5K,
the hysteresis loops become substantially smaller because of the
enhanced thermal effects. In this case, there are deformed step
structures in the lower parts of the magnetization curves and
there does not exists any clear step structure in the upper parts.
The deformed step structures between 1 and 3 T result from the
thermally assisted LZ tunnelings. For $T$=2.5K, there does not
exist any step structure at all for all the six cases. The effect
of the lattice shape is attributed to the long-range property of
the dipolar interaction, and can be clearly seen in the
magnetization curves only at the low temperatures in the extreme
cases of $20\times20\times3$ and $3\times3\times100$. Actually,
there is little visible difference between the magnetization
curves of the three lattices: $12\times12\times8$,
$10\times10\times10$, and $9\times9\times14$. Visible difference
can be found at 0.1K and 1.5K only for the two extreme cases:
$20\times20\times3$ and $3\times3\times100$. If we define a ratio
$r=L_l/L_t$ of longitudinal size to transverse size for $L_t\times
L_t\times L_l$, we have $r$=1 for $10\times10\times10$, $r$=0.67
for $12\times12\times8$, $r$=1.56 for $9\times9\times14$, $r$=0.15
for $20\times20\times3$, and $r$=33 for $3\times3\times100$.
Therefore, there is little clear effect of lattice shape as long
as the shape parameter $r$ is neither extremely large nor
extremely small.

Now we address the statistical errors. We have calculated standard
errors $\sigma_M$ of the reduced magnetization $M/M_s$ as
functions of the sweeping field for various temperatures and
sweeping rates. Our results show that for a given magnetization
curve, the statistical errors are very small ($\sigma_M<0.005$) in
the region of $B_z$ defined by $|M/M_s|>0.9$, and reach a maximal
value $\sigma^{\rm max}_M$ near the point of $B_z$ defined by
$M/M_s=0$. The maximal statistical error $\sigma^{\rm max}_M$ is
dependent on the temperature and sweeping rate, varying from 0.015
to 0.025 for our simulation parameters. Such statistical errors
appear only in a very small region of $B_z$. For any magnetization
curve as a whole, the statistical errors are small enough to be
acceptable.

Here we discuss effects of lattice sizes on simulated results. The
above simulated results are based on the lattice dimensions:
$20\times20\times3$, $12\times12\times8$, $10\times10\times10$,
$9\times9\times14$, and $3\times3\times100$. They have $900\sim
1200$ body-centered unit cells, or $1800 \sim 2400$ spins. To test
our results, we have done a series of simulations for different
parameters using lattice dimension defined by $L_t\times L_t\times
L_l$. In the cases of $T=0.1$K, $\nu=0.2$T/s, and $L_t=L_l=L$ with
$L=5\sim 20$, the largest size effects appear between 4 and 5.5 T
for the right parts of the magnetization curves. For the steps at
4T, the $L$-caused change in the magnetization decreases quickly
with increasing $L$, becoming very small when $L$ is larger than
9. Therefore, our lattice sizes of the results presented above are
large enough to be reliable.

The above simulated results show that the area enclosed by a
magnetization hysteresis loop decreases with the temperature
increasing and increases with the sweeping rate increasing. This
is completely consistent with the temperature and sweeping-rate
dependence of the thermal reversal probability and LZ tunneling
probabilities. Thermal activation effects dominate at high
temperature. The LZ tunneling effects manifest themselves through
the steps and kinks along the magnetization curves. However, there
is a limit for the hysteresis loops at the low temperature end for
a given sweeping rate. These limiting magnetization curves are
caused by the minimal reversal probability set by the direct LZ
quantum tunneling effect because the thermal activation
probability becomes tiny at such low temperatures. With usual
shape parameter $r$, these results are consistent with
experimental magnetization curves of good Mn$_{12}$ crystal
samples in the presence of little misalignments between the easy
axis and applied fields \cite{Mn12tBuAc,TB}. In principle, a
transverse magnetic field (due to the misalignment of the applied
field and the easy axis) can enhance the energy splitting, and as
a result will reduce the magnetization loop and smooth some steps
\cite{Mn12tBuAc,TB,Mn12added}. These usual (not extreme) shape
parameters should reflect real shape factors in experimental
samples. The consistence should be satisfactory, especially
considering that our theoretical probabilities are calculated
under leading order approximation and our model does not include
possible defects and disorders in actual materials.

\begin{table}[!htb]
\caption{Calculated results of $B_{S,m'}^0$, $\Delta B^n_{S,m'}$,
$P_{S,m'}^0$, $\langle P^{\rm LZ,n}_{S,m'}\rangle$, and
$\sigma_{S,m'}^{n}$ for the direct LZ tunneling $(S,m')$ when the
field $B_z$ is swept to 3.75 T, where $n$ is determined by the
field 3.75 T. $T=0.1$ K, $\nu=0.02$ T/s, and the lattice dimension
is $10\times10\times10$. } \label{tab:1}
\begin{ruledtabular}
\begin{tabular}{rccccc}
 $m'$ & $B_{S,m'}^0$(T) & $\Delta B^n_{S,m'}$(T) & $P_{S,m'}^0$ & $\langle P^{\rm LZ,n}_{S,m'}\rangle$ & $\sigma_{S,m'}^{n}$\\ \hline
-10&    0.000000&   6.4$\times10^{-15}$&   0.00000 &  0.00000&   0.00000\\
 -9&    0.564160&   1.6$\times10^{-6~}$&   0       &  0.00000&   0.00000\\
 -8&    1.099966&   3.5$\times10^{-6~}$&   0.00000 &  0.00000&   0.00000\\
 -7&    1.612415&   5.1$\times10^{-6~}$&   0       &  0.00000&   0.00000\\
 -6&    2.106511&   6.7$\times10^{-6~}$&   0.00138 &  0.00138&   0.00001\\
 -5&    2.587260&   7.9$\times10^{-6~}$&   0       &  0.00002&   0.00002\\
 -4&    3.059671&   8.6$\times10^{-6~}$&   0.01815&   0.01838&   0.00320\\
 -3&    3.528757&   8.6$\times10^{-6~}$&   0       &  0.22194&   0.21086\\
 -2&    3.999529&   7.8$\times10^{-6~}$&   1.00000&   1.00000&   0.00000\\
 -1&    4.476997&   6.3$\times10^{-6~}$&   0       &  0.53746&   0.33455\\
  0&    4.966165&   3.9$\times10^{-6~}$&   1.00000&   1.00000&   0.00089\\
  1&    5.472035&   7.4$\times10^{-7~}$&   0       &  0.99975&   0.01091\\
  2&    5.999604&   3.6$\times10^{-6~}$&   1.00000&   1.00000&   0.00000\\
  3&    6.553867&   8.6$\times10^{-6~}$&   0       &  0.99988&   0.00749\\
\end{tabular}
\end{ruledtabular}
\end{table}

\section{Key roles of dipolar fields}

To investigate the effects of dipolar interactions, we divide the
dipolar fields within the $n$-th MC step, $(B_{ix,n}^{\rm di},
B_{iy,n}^{\rm di}, B_{iz,n}^{\rm di})$, into two parts: transverse
dipolar field $B_{ix,n}^{\rm di}$ and $B_{iy,n}^{\rm di}$, and
longitudinal dipolar field $B_{iz,n}^{\rm di}$. Transverse dipolar
field not only modifies $B^{i,n}_{m,m'}$, but also affects
$\Delta^{i,n}_{m,m'}$ and $P^{\rm LZ,i,n}_{m,m'}$. In contrast,
longitudinal dipolar field affects neither $\Delta^{i,n}_{m,m'}$
nor $P^{\rm LZ,i,n}_{m,m'}$, but shifts $B^{i,n}_{m,m'}$ by
$-B_{iz,n}^{\rm di}$. This means that LZ tunnelings actually occur
at the field $B^{i,n}_{m,m'}-B_{iz,n}^{\rm di}$, not
$B^{i,n}_{m,m'}$. This shift has two effects. First, it broadens
the LZ transition and deforms the steps in magnetization curves.
Second, the quick changing of $B_{iz,n}^{\rm di}$ results in that
the value $B^{i,n}_{m,m'}$ can be missed by the effective field
$B_z+B_{iz,n}^{\rm di}$, and therefore the actual percentage of
the reversed spins due to the LZ tunneling effect with respect to
the total spins is smaller than the LZ probability $P^{\rm
LZ,i,n}_{m,m'}$ given in Eq. (\ref{Plz}). This means that the
dipolar interaction hinders both the direct LZ tunneling process
and the thermal assisted LZ tunneling processes.

\begin{figure}[!htbp]
\includegraphics[width=8cm]{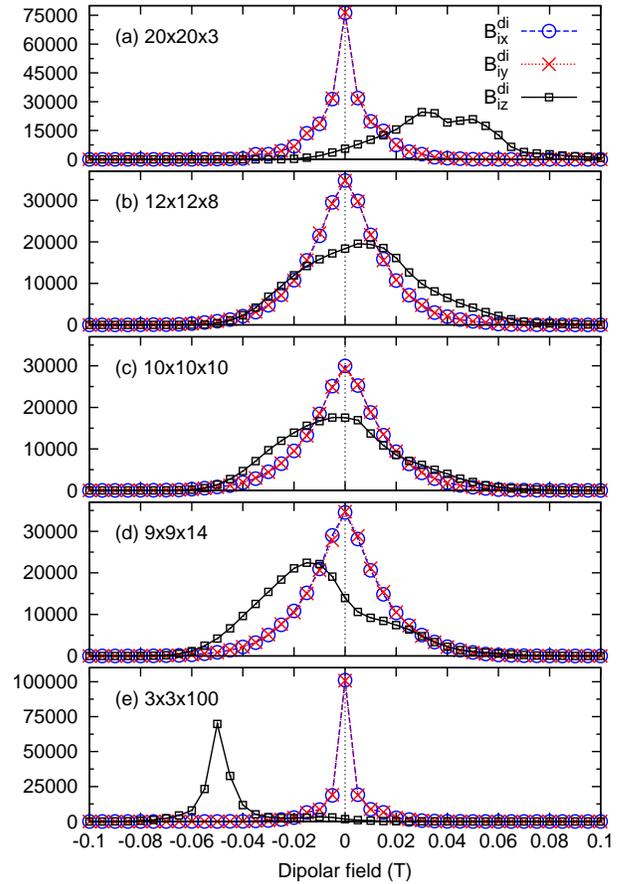}
\caption{(Color online.) Distributions of dipolar fields
$B_{ix}^{\rm di}$ (dashed line + circle), $B_{iy}^{\rm di}$
(dotted line + cross), and $B_{iz}^{\rm di}$ (solid line + square)
for the five lattice dimensions $20\times 20\times 3$ (a),
$12\times12\times8$ (b), $10\times10\times10$ (c), $9\times
9\times14$ (e), and $3\times 3\times100$ (e) when the field $B_z$
is swept to 3.75 T. The temperature $T$ is 0.1 K and the sweeping
rate $\nu$ equals 0.02 T/s. } \label{fig:6}
\end{figure}

Without transverse dipolar field, $B_{m,m'}^{i,n}$ becomes
$B_{m,m'}^0$, and $P^{\rm LZ,i,n}_{S,m'}$ equals 0 for odd $m'$
values because transverse dipolar field is the only transverse
term of odd order in Hamiltonian Eq. (4). Without longitudinal
dipolar field, the V-parts of steps remain vertical and the
percentage of the reversed spins due to LZ tunneling is strictly
equivalent to the LZ probability $P^{\rm LZ,i,n}_{S,m'}$ at low
temperatures. These are shown by the thin solid line for 0.1K in
Fig. 5. In Table I we also present the average value ($\Delta
B^n_{S,m'}$=$\langle | B^{n}_{S,m'}-B_{S,m'}^0|\rangle$) of
dipolar-field fluctuations with respect to $B_{S,m'}^0$, the
dipolar-field-free LZ probability $P^0_{S,m'}$, and the average
value $\langle P^{\rm LZ,n}_{S,m'}\rangle$ and the corresponding
standard error $\sigma_{S,m'}^{n}$ of $P^{\rm LZ,i,n}_{S,m'}$ for
the avoided crossing positions of $S$ and $m'$, where $m'$ varies
from -10 to 3 and the averaging $\langle X^n \rangle$ of $X^{i,n}$
is calculated over all the spins and all the runs within the
$n$-th MC step. It should be pointed out that the $\Delta
B^n_{S,m'}$ values, although very important to LZ tunnelings, are
very small, as shown in Table I. It is transverse dipolar field
that make $\Delta B^n_{S,m'}$ nonzero and make $P^{\rm
LZ,i,n}_{S,m'}$ ($m'$=-5, -3, -1, 1, 3) change from 0 to nonzero,
even nearly reach 1 in the cases $m'=1$ and 3.

In order to elucidate the magnitude and distribution of the
dipolar fields, we address the time-dependent distributions of
SMMs that have dipolar fields $(B_{ix}^{\rm di}, B_{iy}^{\rm di},
B_{iz}^{\rm di})$ (here the continuous time variable is implied),
or in short the distributions of $B_{ix}^{\rm di}$, $B_{iy}^{\rm
di}$, and $B_{iz}^{\rm di}$, in the following. In Fig. 6 we
compare the results from five different lattice dimensions:
$20\times 20\times 3$, $12\times12\times8$, $10\times10\times10$,
$9\times 9\times14$, and $3\times 3\times100$. Here the time is
when the field $B_z$ is swept to 3.75 T, the temperature $T$ is
0.1 K, and the sweeping rate $\nu$ equals 0.02 T/s. For all the
five lattices, our results show that the distribution of
$B_{ix}^{\rm di}$ always is approximately equivalent to that of
$B_{iy}^{\rm di}$ and they are both symmetrical and peaked at
zero. The peak is sharper for the extremely slab-like
$20\times20\times3$ lattice and extremely rod-like
$3\times3\times100$ lattice. The peak of the $B_{iz}^{\rm di}$
distribution is wider than that of both $B_{ix}^{\rm di}$ and
$B_{iy}^{\rm di}$. It shifts substantially away from zero when the
lattice shape is either extremely slab-like or extremely rod-like.
The leftward shift of the $B_{iz}^{\rm di}$ peak can be attributed
to dipolar-interaction-induced ferromagnetic orders in rod-like
systems \cite{garanin1,garanin2}, and the similar rightward shift
to dipolar-interaction-induced antiferromagnetic orders in
slab-like systems. Because dipolar interactions are the only
inter-SMM interactions in our model, the differences of
distributions between the the five lattices are caused by the
dipolar fields, or dipolar interactions in essence.

\section{conclusion}

In summary, we have combined the thermal effects with the LZ quantum
tunneling effects in a DMC framework by using the giant spin
approximation for spins of SMMs and considering magnetic dipolar
interactions for comparison with experimental results. We consider
ideal lattices of SMMs consistent with experimental ones and assume
that there are no defects and axis-misalignments therein. We
calculate spin reversal probabilities from thermal-activated barrier
hurdling, direct LZ tunneling effect, and thermal-assisted LZ
tunneling effects in the presence of sweeping magnetic fields.
Taking the parameters of experimental Mn$_{12}$ crystals, we do
systematical DMC simulations with various temperatures and sweeping
rates. Our results show that the step structures can be clearly seen
in the low-temperature magnetization curves, the thermally activated
barrier hurdling becomes dominating at high temperature near 3K, and
the thermal-assisted tunneling effects play important roles at the
intermediate temperature. These are consistent with corresponding
experimental results on good Mn$_{12}$ samples (with less disorders)
in the presence of little misalignments between the easy axis and
applied fields \cite{Mn12tBuAc,TB}, and therefore our magnetization
curves are satisfactory.

Furthermore, our DMC results show that the magnetic dipolar
interactions, with the thermal effects, have important effects on
the LZ magnetization tunneling effects. Their longitudinal parts
can partially break the resonance conditions of the LZ tunnelings
and their transverse parts can modify the tunneling probabilities.
They can clearly manifest themselves when the SMM crystal is
extremely rod-like or slab-like. However, both the magnetic
dipolar interactions and the LZ tunneling effects have little
effects on the magnetization curves when the temperature is near
3K. This DMC approach can be applicable to other SMM systems, and
could be used to study other properties of SMM systems.

\begin{acknowledgments}
This work is supported  by Nature Science Foundation of China
(Grant Nos. 10874232 and 10774180), by the Chinese Academy of
Sciences (Grant No. KJCX2.YW.W09-5), and by Chinese Department of
Science and Technology (Grant No. 2005CB623602).
\end{acknowledgments}


\end{document}